\documentclass[aps,pra,twocolumn,groupedaddress,showpacs]{revtex4}

\usepackage{bm}
\usepackage{amssymb}
\usepackage{graphicx,epsfig}

\begin{document}
\title{ Coupling qubits in circuit-QED cavities connected by a bridge qubit}
\author{Mun Dae Kim and Jaewan Kim}
\affiliation{ Korea Institute for Advanced Study, Seoul 130-722, Korea}
\date{\today}

\begin{abstract}
We analyze a coupling scheme for qubits in different cavities of
circuit-QED architecture. In contrast to the usual scheme where
the cavities are coupled by an interface capacitance we employ a
bridge qubit connecting cavities to mediate two-qubit coupling.
This active coupling scheme makes it possible to switch on/off and
adjust the strength of qubit-qubit coupling, which is essential
for scalability of quantum circuit. By transforming the
Hamiltonian we obtain an exact expression of two-qubit coupling in
the rotating-wave approximation. For the general case of $n$
qubits  the Hamiltonian can produce the W state as an eigenstate
of the system. We calculate the decay rate of the coupled
qubit-resonator system to find that it is viable in real
experiments.
\end{abstract}

\maketitle

\section{Introduction}
In the circuit quantum electrodynamics (QED)  architecture the
oscillating modes in superconducting transmission line resonator
interact with superconducting qubits coupled with the resonator.
The  circuit-QED architecture has the advantage that the dipole
moment of the qubit (an artificial atom) can be adjusted up to a
significantly large value compared to that of atom in the
cavity-QED, which gives rise to a strong qubit-resonator coupling
\cite{Blais1,Blais}.
%
Two qubits coupled with the same resonator can interact with each
other through resonator modes \cite{Majer,DiCarlo,Loo,Kim}.
Recently three-qubit quantum error correction code has been
implemented in a circuit-QED device where four qubits are coupled
with the same resonator \cite{Reed}.

In order to couple more qubits one can employ Jaynes-Cummings type
lattice structure consisting of circuit-QED circuits, where each
transmission line resonator is coupled to just one qubit and the
interaction between resonators is mediated via a coupling
capacitance \cite{Schmidt,Houck,Underwood,Leib,Gerace,Tian}.
Further, the coupling capacitance can be replaced by a Josephson ring consisting of
superconducting ring interrupted by  Josephson
junctions in order to demonstrate the effect of time-reversal
symmetry breaking \cite{Koch,Nunnenkamp}.
In these {\it passive} coupling the coupling circuit only transmits photons to the other
resonator through virtual excitations.

On the other hand, the cavity modes can be coupled via an
interface qubit instead  of the coupling capacitance
\cite{Strauch10,Strauch,Sharma}. In a recent experiment two
cavities in a three-dimensional circuit-QED architecture has been
coupled through a {\it bridge qubit} \cite{Paik}. In this study,
we consider  a circuit-QED system consisting of two qubits coupled
to  two cavities which are connected through  a two-level system
(a bridge qubit) as shown in Fig. \ref{fig1}(a). While  previous
studies couples the photon modes in cavities without qubit
\cite{Koch,Nunnenkamp,Paik}, we introduce a qubit in each cavity
and couple these qubits by using the bridge qubit. In this {\it
active} coupling the bridge qubit interacts with the photon modes
of the cavities, and thus by controlling the bridge qubit state we
can switch on/off and adjust the strength of the qubit-qubit
coupling.

The transformed Hamiltonian can be represented as a direct sum of
two Hamiltonians: the one describes the resonator modes and the
other the qubit and bridge qubit states. This transformation
provides an exact expression of two-qubit coupling in latter
Hamiltonian.
If one introduce the transmon qubit as a bridge qubit, one can
control the two-qubit coupling by  controlling the transmon state,
which is essential for scalability of quantum circuit.
More than two qubits can be coupled through a bridge qubit,
which may realize the circuit-QED lattice model \cite{Schmidt,Houck,Underwood,Koch,Nunnenkamp}
with controllable interaction.
We consider the general case of $n$ qubits interacting via a bridge
qubit as shown in Fig. \ref{fig1}(b). We show that the $n$-qubit
Hamiltonian can be transformed to an effective model with
xy-type  coupling and the W state can be formed as an eigenstate of the system.


For two qubit case we provide numerical results for the two-qubit
coupling and discuss achieving two-qubit coupling and $\sqrt{{\rm
iSWAP}}$ gate  in experiments. Further, we obtain the eigenstates
of the coupled qubit-resonator system and calculate the decay rate
of the  system. The decay rate is of the same order of that of the
uncoupled qubit state so that the coupled qubit-resonator system
may be viable in real experiments.

\section{Hamiltonian of coupled qubits}

\begin{figure}[b]
\vspace{-0.3cm}
\includegraphics[width=0.6\textwidth]{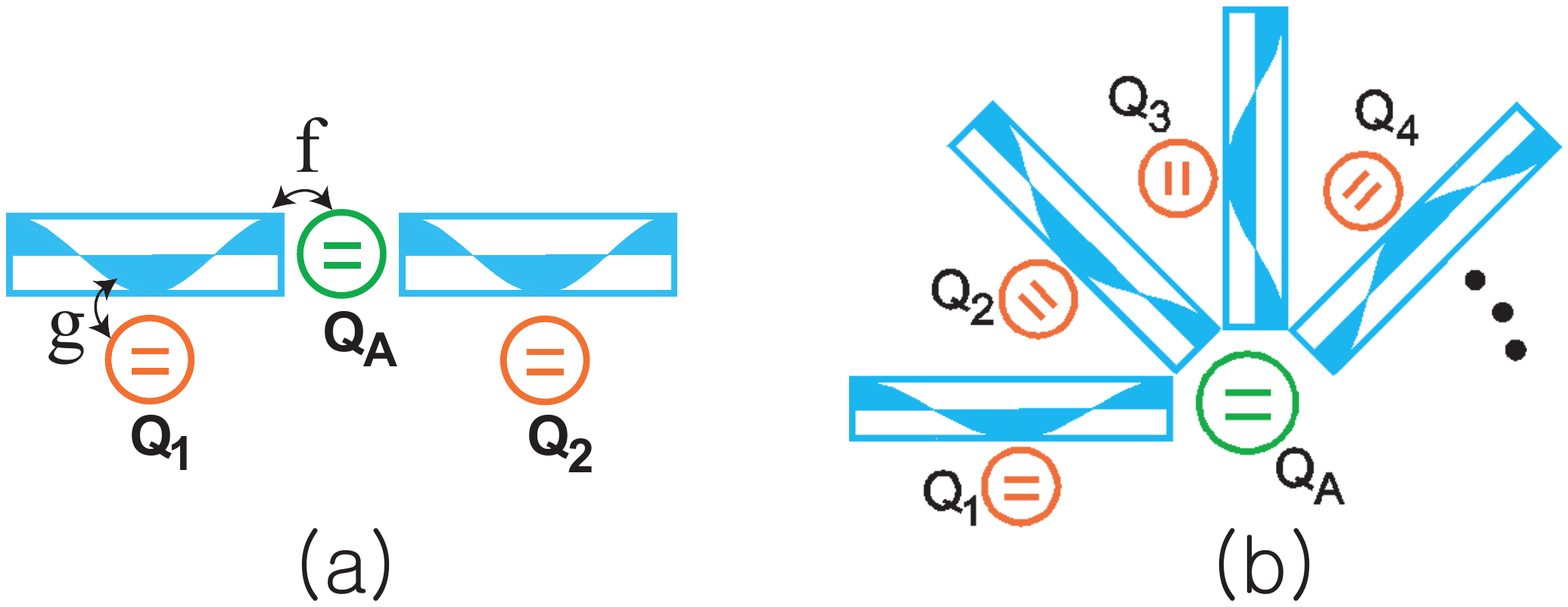}
\vspace{-4.5cm}
\caption{(a) Two qubits $Q_1$ and $Q_2$ coupled via a bridge
qubit $Q_A$ in circuit-QED architecture.  The coupling between
qubit (bridge qubit) and the resonator is $g$ ($f$) at the
anti-node of the resonating modes. (b) $n$  qubits coupled via a bridge qubit. }
\label{fig1}
\end{figure}

Figure \ref{fig1}(a) shows two transmission
line resonators of circuit-QED scheme coupled to two qubits ($Q_1$ and $Q_2$) and a
two-level system ($Q_A$, a bridge qubit) at the antinodes of the
resonator modes.  The bridge qubit is coupled at the end of
resonator as shown in the circuit-QED scheme of Ref.
\onlinecite{Blais}. The resonating modes in the resonators
interact with the two-level system as well as the qubits, resulting in virtual qubit-qubit
coupling. In general, more resonators can be coupled to the
two-level system as shown in Fig. \ref{fig1}(b), and thus we will
analyze the Hamiltonian for general $n$ qubits coupled via a bridge
qubit.

The $n$-qubit Hamiltonian can be written in the rotating-wave approximation as
\begin{eqnarray}
\label{H}
H_{n}&=&\frac{1}{2}\omega_a\sigma_{az}
+\sum^n_{j=1}\left[\omega_{rj} a^\dagger_j a_j+\frac12 \omega_{qj}\sigma_{jz} \right.\\
&&\left. -g_j(a^\dagger_j\sigma_{j-}+a_j\sigma_{j+})-f_j(a^\dagger_j\sigma_{a-}+a_j\sigma_{a+})\right], \nonumber
\end{eqnarray}
where $\omega_a, \omega_{qj}$ and $\omega_{rj}$ are the frequencies of
two-level system $Q_A$, qubit $Q_j$, and corresponding resonating mode, respectively.
The qubit $Q_j$ and  resonating mode are coupled  with the coupling constant $g_j$
at the center of the resonator, and
$f_j$ is the coupling constant between the bridge qubit and the resonating mode
at the end of the resonator.

The off-diagonal terms with the coupling constants $g_j$ and $f_j$
can  be eliminated by introducing the transformation,
\begin{eqnarray}
\label{tilH}
{\tilde H}_{n}=U^\dagger_{n} H_{n} U_{n},
\end{eqnarray}
where
\begin{eqnarray}
\label{Un}
U_{n}=e^{M}=e^{-\sum^n_{j=1}[\phi_j(a^\dagger_j\sigma_{j-}-a_j\sigma_{j+})
+\theta_j(a^\dagger_j\sigma_{a-}-a_j\sigma_{a+})]}.
\end{eqnarray}
We, for simplicity, consider identical qubits and resonators, and
set  $\omega_{rj}=\omega_r$, $\omega_{qj}= \omega_q$,
$\phi_j=\phi$, and $\theta_j=\theta$.
The Hamiltonian of Eq. (\ref{H}) conserves the excitation number,
\begin{eqnarray}
{\cal N}_e=\sum^n_{j=1}(s_{jz}+1/2+N_{rj})+(s_{az}+1/2),
\end{eqnarray}
where  $s_{jz}, s_{az} \in\{-1/2,1/2\}$ with $j\in\{1,\cdots,n\}$
are the eigenvalues of the operators $S_{jz}=\frac12\sigma_{jz}$ and $S_{az}=\frac12\sigma_{az}$,
respectively, and $N_{rj}$ is the excitation number of oscillating modes in
$j$-th resonator. Here, we consider the lowest excitation case
that ${\cal N}_e=1$ and thus $N_{rj}\in\{0,1\}$.

$U_{n}$ is a block-partitioned $(2n+1)\times(2n+1)$ matrix in the basis of
$\{|s_{1z},N_{r1},s_{2z},N_{r2},$ $\cdots ,s_{nz},N_{rn},s_{az}\rangle\}$.
The explicit form of $U_{n}$ for  ${\cal N}_e=1$ can be calculated as
\begin{eqnarray}
\label{MUn}
U_{n}=\left(
\begin{array}{cccccc}
Q & T & T & \cdots & T & C \\
T & Q & T & \cdots & T & C \\
T & T & Q & \cdots & T & C \\
\vdots & \vdots & \vdots & \ddots &\vdots &\vdots \\
T & T & T & \cdots & Q & C \\
{C'} & C' & C' & \cdots& C'  & \frac{\phi^2+n\theta^2\cos q}{q^2} \\
\end{array}
\right),
\end{eqnarray}
where the matrices $Q, T, C,$ and $C'$  are derived in Appendix.

In the transformed Hamiltonian of Eq. (\ref{tilH})
the coupling terms in Eq. (\ref{H}) can be eliminated by introducing
the conditions,
\begin{eqnarray}
\label{c1}
\tan 2\phi&=&2g/\Delta,\\
\label{c2}
\tan 2q&=&\frac{2q(\phi g+n\theta f)}{q^2\Delta+n\theta^2\Delta'},\\
\label{c3}
\tan q&=&\frac{q(\phi f-\theta g)}{\theta\phi\Delta'}
\end{eqnarray}
with  $\Delta\equiv \omega_q-\omega_r$, $q\equiv
\sqrt{\phi^2+n\theta^2}$,  and $\Delta'\equiv \omega_a-\omega_q$,
which ensures $[{\tilde H}_{n}]_{uv}=0$ if $u+v$=odd. For given
$g/\Delta$ and $f/\Delta$, the variables $\phi$, $\theta$, and
$\Delta'/\Delta$ are determined from these conditions.

As a result, ${\tilde H}_{n}$ can be represented as  a block-diagonal form,
\begin{eqnarray}
{\tilde H}_{n}= {\tilde H}_{n,1}\oplus {\tilde H}_{n,2}.
\end{eqnarray}
Here ${\tilde H}_{n,1}$ is an $n\times n$ matrix whose basis is
given by the transformation   of
$\{|-1/2,N_{r1},-1/2,N_{r2},\cdots ,-1/2,$ $ N_{rn},-1/2\rangle\}$
by $U_{n}$, describing the resonator modes: one of the resonator
modes is excited while all the qubits and bridge qubit are at the
ground states. $ {\tilde H}_{n,2}$ is written by an  $(n+1)\times
(n+1)$ matrix with the basis obtained by the same transformation
of $\{|s_{1z},0,s_{2z},0,$ $\cdots ,s_{nz},0,s_{az}\rangle\}$ such
as
\begin{eqnarray}
{\tilde H}_{n,2}=\left(
\begin{array}{cccccc}
e_q & j_q & j_q & \cdots & j_q & j_a \\
j_q & e_q & j_q & \cdots & j_q & j_a \\
j_q & j_q & e_q & \cdots & j_q & j_a \\
\vdots & \vdots & \vdots & \ddots &\vdots &\vdots \\
j_q & j_q & j_q & \cdots & e_q & j_a \\
j_a & j_a & j_a & \cdots& j_a  & e_a \\
\end{array}
\right),
\end{eqnarray}
describing the qubit states: one of the qubits or bridge qubit is
excited while there is no resonator mode excitation.

The interaction between qubits is described by the Hamiltonian  $
{\tilde H}_{n,2}$. The energy levels of the system of $n$-qubits
with a bridge qubit are as follows: for the state with
$s_{jz}=1/2$ for only one qubit, $s_{jz}=-1/2$ for other qubits,
and $s_{az}=-1/2$, $[{\tilde H}_{n,2}]_{ii}=e_q$, and for the
state with  $s_{jz}=-1/2$ for all qubits and $s_{az}=1/2$,
$[{\tilde H}_{n,2}]_{n+1,n+1}=e_a$, where $1\leq i,j \leq n$. The
qubit-qubit coupling is given by $[{\tilde H}_{n,2}]_{ij}=j_q$ for
$i\neq j$ and the coupling between a qubit and the bridge qubit by
$[{\tilde H}_{n,2}]_{i,n+1}=[{\tilde H}_{n,2}]_{n+1,j}=j_a$. Then,
$e_q$, $e_a$, $j_q$, and $j_a$ can be explicitly evaluated as
\begin{eqnarray}
j_q\!\!\!\!&=&\!\!\!-\!\frac{1}{2nq^2}\!(q^2\!\!A\!\!-\!\!\phi^2\!B\!\!-\!\!n\theta^2\!\!\Delta\!)
\!\!+\!\!\frac{\Delta'\!\phi^2\!\theta^2\!\!}{q^4}(\!1\!\!-\!\!2\!\sec q\!\!+\!\cos^2\!\!q),\\
j_a\!\!\!&=&\!\!\!\frac{\phi\theta}{2q^2}(B\!\!-\!\!\Delta)\!\!
+\!\!\frac{\Delta'\phi\theta}{q^4}\![n\theta^2\!\!\cos^2\!\! q\!\!+\!\!(\phi^2\!\!\!-\!\!n\theta^2)\!\sec q\!-\!\!\phi^2],\\
e_q\!\!\!&=&\!\!\!\frac{n-1}{2n}A+\frac{\phi^2 B-n(q^2-\theta^2)\Delta}{2nq^2}\\
&&+\frac{\Delta'\phi^2\theta^2}{q^4}(1\!-\!2\sec q\!+\!\cos^2q)
\!-\!\frac{\omega_a}{2}\!+\!\left(1\!-\!\frac{n}{2}\right)\!\omega_q,\nonumber\\
e_a\!\!\!&=&\!\!\!\frac{n^2\theta^2}{2nq^2}(B\!\!-\!\!\Delta)
\!\!+\!\!\frac{\Delta'}{q^4}(\phi^4\!\!+2n\phi^2\theta^2\sec q+\!n^2\theta^4\!\cos^2 q)\nonumber\\
&&\!\!\!-\!\frac{\omega_a}{2}\!+\!\!\left(\!1\!-\!\frac{n}{2}\!\right)\!\omega_q,
\end{eqnarray}
where $A\equiv\Delta\cos2\phi+2g\sin2\phi$ and
$B\equiv\Delta\cos2q+(2/q)(n\theta f+\phi g)\sin 2q$.

Further, the coupling term $j_a$ between a qubit and the bridge qubit can be eliminated by the transformation,
\begin{eqnarray}
\label{Hstar}
H^*_{n,2}={\tilde U}^\dagger_{n,2} {\tilde H}_{n,2} {\tilde U}_{n,2},
\end{eqnarray}
where
\begin{eqnarray}
{\tilde U}_{n,2}=e^{\tilde M}=e^{-\sum^n_{j=1}\eta_j(\sigma_{j+}\sigma_{a-} - \sigma_{j-}\sigma_{a+})}.
\end{eqnarray}
For the identical qubit case, $\eta_j=\eta$,
${\tilde U_{n,2}}$ can be explicitly evaluated as before,
\begin{eqnarray}
{\tilde U}_{n,2}\!\!=\!\!\!\left(
\begin{array}{ccccc}
\!\!\frac{n-1+\cos\sqrt{n}\eta}{n} & \!\!\frac{\cos\sqrt{n}\eta-1}{n} & \!\! \cdots & \!\!\frac{\cos\sqrt{n}\eta-1}{n}
& \!\!\!-\frac{\sin\sqrt{n}\eta}{\sqrt{n}} \\
\!\!\frac{\cos\sqrt{n}\eta-1}{n} & \!\!\!\!\!\frac{n-1+\cos\sqrt{n}\eta}{n} & \!\! \cdots & \!\!\frac{\cos\sqrt{n}\eta-1}{n}
& \!\!\!-\frac{\sin\sqrt{n}\eta}{\sqrt{n}} \\
\!\!\vdots & \vdots & \!\!\ddots & \vdots & \!\!\!\vdots \\
\!\!\frac{\cos\sqrt{n}\eta-1}{n} & \!\!\frac{\cos\sqrt{n}\eta-1}{n} & \!\! \cdots & \!\!\!\!\frac{n-1+\cos\sqrt{n}\eta}{n}
& \!-\!\frac{\sin\!\!\sqrt{n}\eta}{\sqrt{n}} \\
\!\!\frac{\sin\sqrt{n}\eta}{\sqrt{n}} & \!\!\frac{\sin\sqrt{n}\eta}{\sqrt{n}} & \!\!\cdots
& \!\!\frac{\sin\sqrt{n}\eta}{\sqrt{n}}  & \!\!\!\cos\sqrt{n}\eta \\
\end{array}
\!\!\right)\!\!.\nonumber\\
\end{eqnarray}


By introducing the condition with $\epsilon\equiv e_a-e_q$,
\begin{eqnarray}
\label{tan} \tan
2\sqrt{n}\eta=\frac{2\sqrt{n}j_a}{(n-1)j_q-\epsilon},
\end{eqnarray}
determining the variable $\eta$, we have
\begin{eqnarray}
\label{Hn2*}
H^*_{n,2}=\left(
\begin{array}{cccccc}
\varepsilon^q_n & J_{n} & J_{n} & \cdots & J_{n} &0 \\
J_{n} & \varepsilon^q_n & J_{n} & \cdots & J_{n} &0 \\
J_{n} &  J_{n} & \varepsilon^q_n & \cdots & J_{n} &0 \\
\vdots & \vdots & \vdots & \ddots   &\vdots  &\vdots \\
J_{n} &  J_{n}  & J_{n} &\cdots &  \varepsilon^q_n &0 \\
0 & 0 & 0& \cdots & 0 & \varepsilon^a_n
\end{array}
\right).
\end{eqnarray}
Here the energy levels are given by
\begin{eqnarray}
\label{eqn}
\varepsilon^q_n\!\!\!\!&=&\!\!\!\frac{1}{n}(\epsilon\!-\!(n\!-\!1)j_q)\sin^2\!\!\!\sqrt{n}\eta\!
\!+\!\!\frac{1}{\sqrt{n}}j_a\sin2\!\sqrt{n}\eta\!+\!e_q,\\
\label{eqa}
\varepsilon^a_n\!\!\!\!&=&\!\!\!\epsilon\!\cos^2\!\!\!\!\sqrt{n}\eta\!\!+\!(n\!\!-\!\!1\!)j_q\!
\sin^2\!\!\!\!\sqrt{n}\eta\!-\!\!\sqrt{n}j_a\!\sin\!2\!\sqrt{n}\eta\!\!+\!e_q,
\end{eqnarray}
and the qubit-qubit coupling
\begin{eqnarray}
\label{Jn}
J_{n}\!\!&=&\!\!\frac{1}{2n}[(n+1)j_q+\epsilon+((n-1)j_q-\epsilon)\cos 2\sqrt{n}\eta \nonumber\\
&&+2\sqrt{n}j_a\sin2\sqrt{n}\eta].
\end{eqnarray}
This qubit-qubit coupling strength   can be
explicitly written as
\begin{eqnarray}
\label{Jn}
J_{n}\!=\!\frac{1}{2n}\!\left[(n+1)j_q\!+\!\epsilon+\!\sqrt{((n-1)j_q\!-\!\epsilon)^2\!+\!4nj_a^2}
\right]
\end{eqnarray}
by using the condition of Eq. (\ref{tan}).
Here we discard a physically meaningless solution which has a finite value even for $f=0$ or $g=0$.

\begin{figure}[b]
\vspace{-0cm}
\includegraphics[width=0.65\textwidth]{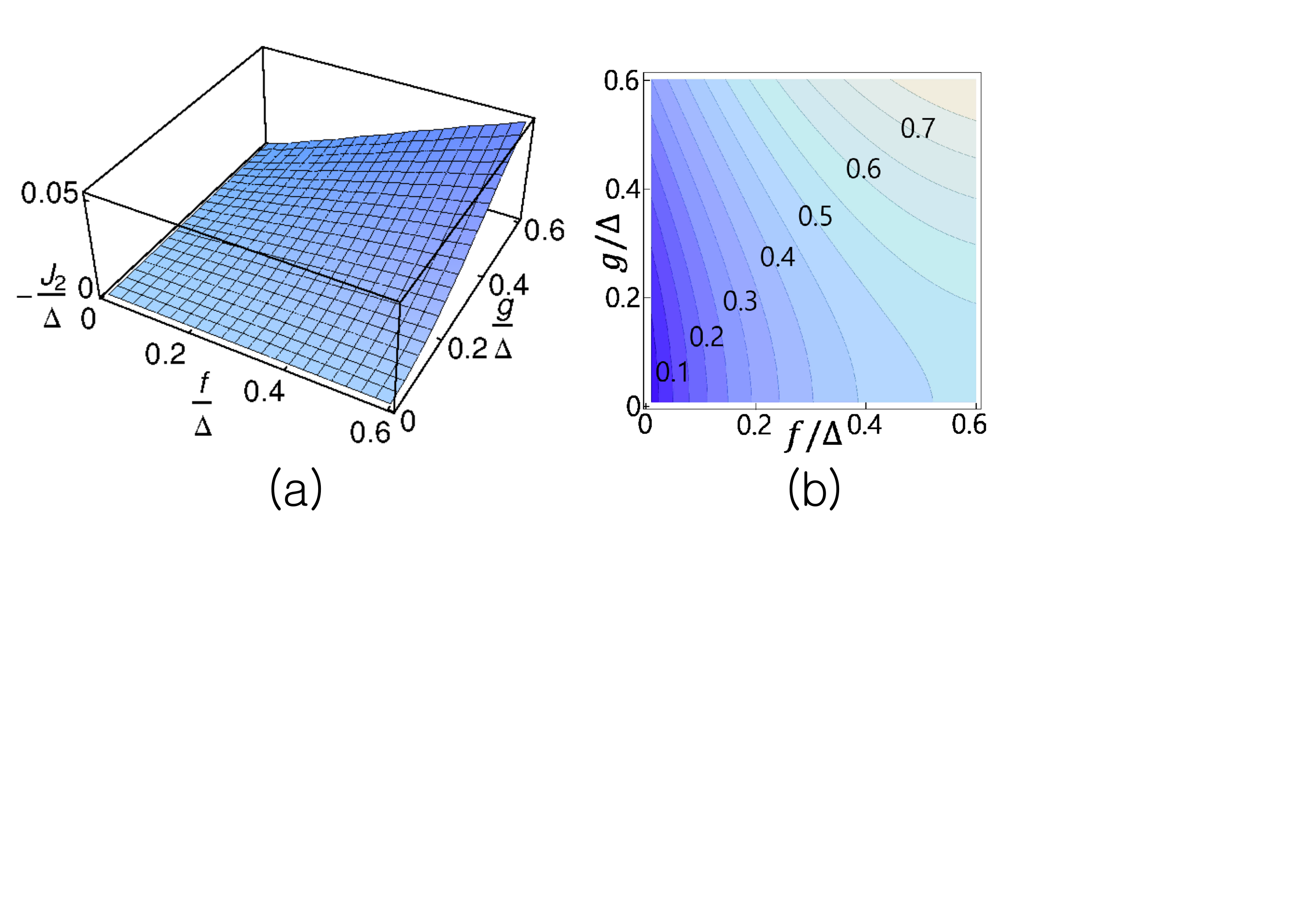}
\vspace{-4cm}
\vspace{0cm}
\caption{(a) Two qubit  xy-coupling  $J_{2}$ as a function of
$g/\Delta$ and $f/\Delta$. (b) A contour plot for
$\Delta'/\Delta$ obtained from Eqs. (\ref{c1})-(\ref{c3}) for $n=2$ in the
$(g/\Delta,f/\Delta)$ plane.}
\label{fig2}
\end{figure}

The Hamiltonian in Eq. (\ref{Hn2*}) can be represented in the subspace of the Hilbert space
satisfying ${\cal N}_e=1$ as follows:
\begin{eqnarray}
\label{H*n2} H^*_{n,2}\!=\!\frac12\omega'_a\sigma_{az}\!\!+\!\sum^{n}_{j=1}\frac12\omega'_q\sigma_{jz}\!
+\!\!\!\!\sum^n_{i,j=1,i\neq j}
\!\!\!\!\!\!J_{n}(\sigma_{i+}\sigma_{j-}\!\!+\!\sigma_{i-}\sigma_{j+}).\nonumber\\
\end{eqnarray}
For the state with $s_{az}=-1/2$ and one of $s_{iz}$'s is 1/2
we have $\varepsilon^q_n=-\frac{(n-2)}{2}\omega'_q-\frac{1}{2}\omega'_a$, and
for the state with $s_{az}=1/2$ and all $s_{iz}$'s are -1/2 we have $\varepsilon^a_n=-\frac{n}{2}\omega'_q+\frac{1}{2}\omega'_a$,
resulting in the relations
\begin{eqnarray}
\omega'_a&=&-\frac{1}{n-1}(n\varepsilon^q_n-(n-2)\varepsilon^a_n),\\
\label{omega-prime-q}
\omega'_q&=&-\frac{1}{n-1}(\varepsilon^q_n+\varepsilon^a_n).
\end{eqnarray}
Since the Hamiltonian $H^*_{n,2}$ with xy-type coupling has the W state as an eigenstate,
we can produce the W state in the transformed coordinate with the system of
$n$-qubits coupled with each other as shown in Fig. \ref{fig1}(b).


%

\begin{figure}[b]
\vspace{-0cm}
\includegraphics[width=0.8\textwidth]{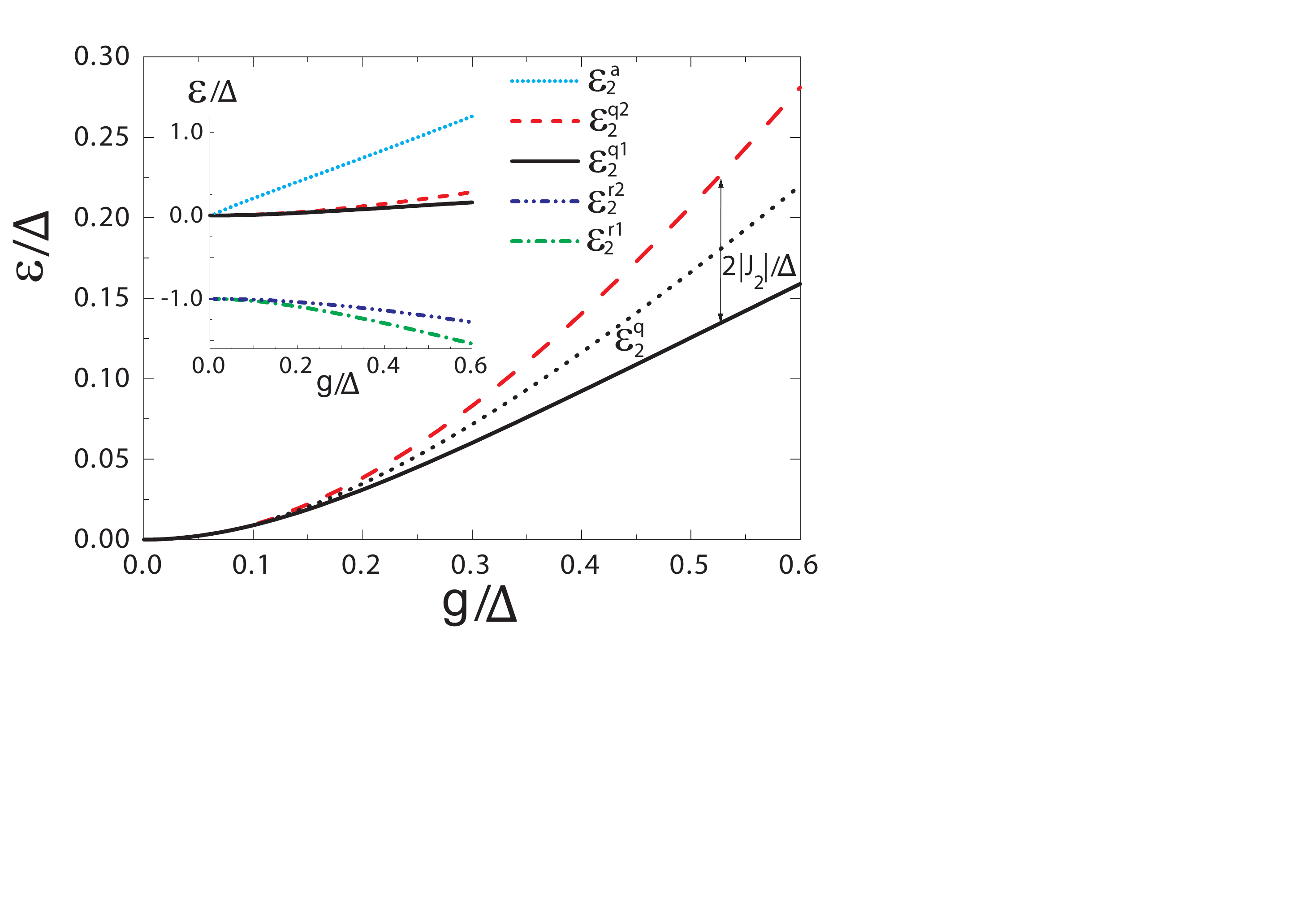}
\vspace{-3.5cm}
\caption{Energy levels of the Hamiltonian
${H}_{2}$ and the two-qubit coupling
strength $J_{2}$ for $g=f$. The dotted line shows the energy
level, $\varepsilon^q_2$. In the inset all the energy levels of
$H_{2}$ are shown, among which the energy levels,
$\varepsilon^{q1}_2=\varepsilon^q_2-|J_{2}|$ and
$\varepsilon^{q2}_2=\varepsilon^q_2+|J_{2}|$, are enlarged in the main
figure.} \label{fig3} \vspace{0cm}
\end{figure}

\section{Two-qubit coupling}

In this section we will give numerical results of two qubit case ($n=2$).
If we set $\omega_{r,1}=\omega_{r,2}=\omega_r$ and
$\omega_{q,1}=\omega_{q,2}=\omega_q$ as before, the transformed
Hamiltonian ${\tilde H}_{2}=U^\dagger_{2} H_{2} U_{2}$
becomes block-diagonalized as
\begin{eqnarray}
{\tilde H}_{2}= {\tilde H}_{2,1}\oplus {\tilde H}_{2,2}
\end{eqnarray}
and we can determine $\Delta'/\Delta$, $\phi$, and $\theta$
by using the conditions in Eqs. (\ref{c1})-(\ref{c3}).
As a  result, we have
\begin{eqnarray}
{\tilde H}_{2,2}=\left(
\begin{array}{ccc}
e_q & j_q & j_a \\
j_q & e_q & j_a \\
j_a & j_a & e_a
\end{array}
\right),
\end{eqnarray}
where the basis is given by the transformation of
$\{|1/2,0,-1/2,0,-1/2\rangle, |-1/2,0,1/2,0,-1/2\rangle, |-1/2,0,-1/2,0,1/2\rangle\}$
by $U_{2}$.
Further, $j_a$ term can be eliminated by the transformation
$H^*_{2,2}={\tilde U}^\dagger_{2,2} {\tilde H}_{2,2} {\tilde U}_{2,2}$, and we have
\begin{eqnarray}
\label{H22*}
H^*_{2,2}=\left(
\begin{array}{ccc}
\varepsilon^q_2 & J_2 & 0 \\
J_2 & \varepsilon^q_2 & 0 \\
0 & 0 & \varepsilon^a_2
\end{array}
\right)
\end{eqnarray}
with $\varepsilon^q_2$, $\varepsilon^a_2$, and $J_2$  in Eqs. (\ref{eqn})-(\ref{Jn}).
%


In Fig. \ref{fig2}(a) we show  $J_{2}$ as a function of
$g/\Delta$ and $f/\Delta$,  where $J_{2}$ increases monotonically
from 0 as $g/\Delta$ or $f/\Delta$ increases. Fig. \ref{fig2}(b)
shows $\Delta'/\Delta$  for obtaining the final Hamiltonian of Eq.
(\ref{H22*}), which requires that we adjust  the value
$\Delta'/\Delta$  as $g$ or $f$ varies by changing the bridge qubit
frequency.
If we consider the bridge qubit as a transmon, this can be done
by varying the parameter value  $E_J/E_C$ with $E_J$ being the
Josephson energy and $E_C$ the charging energy. In the design of
transmon \cite{Koch2} the Josephson energy is given by
$E_J=E_{J,{\rm max}}|\cos(\pi\Phi/\Phi_0)|$ which can be tuned by
the external magnetic flux $\Phi$.
From the numerical data of Ref. \onlinecite{Koch2}
$\delta\omega_a$ is tunable such that $\delta\omega_a \sim $1GHz
for $\delta (E_J/E_C)\sim 10$ around  $E_J/E_C =50$ for a typical
transmon \cite{Schreier}. In Fig. \ref{fig2}(b) $\Delta'/\Delta <
1$,  which can be obtained by adjusting  $\delta \Delta'\sim
\delta \omega_a \lesssim$1GHz with $\Delta\approx $1GHz
 for a circuit-QED architecture \cite{Blais}.

In the inset of Fig. \ref{fig3} the eigenvalues of the two-qubit
Hamiltonian $H_{2}$ are shown for the case that $g=f$.
In the figure $\varepsilon^{r1}_2$ and $\varepsilon^{r2}_2$
correspond to the energy levels of the resonator modes,
and $\varepsilon^{q1}_2$, $\varepsilon^{q2}_2$, and $\varepsilon^{a}_2$
to those of two qubits and bridge qubit, respectively. In Fig. \ref{fig3} we
enlarge the energy levels in the inset corresponding to
$\varepsilon^{q1}_2$ and $\varepsilon^{q2}_2$.
Under the unitary transformations, $U_{2}$ and ${\tilde U}_{2,2}$, the
eigenvalues are invariant, and thus ${\tilde H}_{2,1} \oplus
H^*_{2,2}$ has the same eigenvalues as $H_{2}$.
Since from the Hamiltonian $H^*_{2,2}$ in Eq. (\ref{H22*})
$\varepsilon^{q1}_2=\varepsilon^q_2-|J_{2}|$ and
$\varepsilon^{q2}_2=\varepsilon^q_2+|J_{2}|$,
we can observe that the energy gap in Fig. \ref{fig3}
is the two-qubit coupling strength which is
consistent with the result in Fig. \ref{fig2}(a).

Consider the bridge qubit as the states of transmon at degeneracy point.
Then, far from the degeneracy point of the bridge qubit two-qubits can be
effectively  decoupled, which provides a switching function.  In
Ref.  \onlinecite{Majer} a Stark pulse is applied bringing qubits
in resonance for a quarter period of oscillation to achieve the $\sqrt{{\rm
iSWAP}}$ gate \cite{McDermott}. In this case two qubits  are coupled with the same
resonator with the interaction strength $J=g_1g_2/\Delta$ for the
same qubit frequencies, which is derived from the rotating wave
approximation and the transformation similar to that in Eq.
(\ref{Un}) \cite{Blais}. In the present case we perform the
$\sqrt{{\rm iSWAP}}$ gate similarly by using the switching
function: a Stark pulse bring  the bridge qubit to degeneracy
point and stay for an adjusted time during which two-qubit state evolves.
At the end of the  $\sqrt{{\rm iSWAP}}$ gate the bridge qubit
moves away from the degeneracy point and two-qubit coupling is switched off.

In addition we can consider an one-dimensional qubit array. Owing to
the switching function we can selectively choose a pair of nearest
neighbor qubits and perform a two-quibt gate, and do
the same process for another pair of qubits. In this way we can
implement a scalable quantum computing in 1D array of qubits.
Moreover, since the qubit-resonator couplings $g$ and $f$ can be
tuned by using a three island version of the transmon
\cite{Gambetta}, the two-qubit coupling $J_{n}$ in Eq. (\ref{Jn})
can also be  {\it in situ} controllable. The switching function
and controllability is important for scalability of quantum circuit.

\section{Discussions and Summary}

In section 2 we showed that the W state $|\psi'^*_W\rangle$  can
be formed with the coupled qubits in Fig. \ref{fig1}(b) from the
Hamiltonian in Eq. (\ref{H*n2}) in the transformed coordinate.
However, this is not the exact W state in the original basis.
For the transformed Hamiltonian $H^*=U^\dagger H U$ with the eigenstate such that
$H^*|\psi'^*_W\rangle=E|\psi'^*_W\rangle$  the eigenstate $|\psi'_W\rangle$ of original
Hamiltonian $H$ can be calculated as $|\psi'_W\rangle=U|\psi'^*_W\rangle$.

Actually we have introduced two consecutive transformation, $U_2$
and ${\tilde U}_{2,2}$, and thus we can obtain $|\psi'_W\rangle$
represented as $|\psi'_W\rangle
=\sum^n_{k=1}(c_{2k-1}|\phi_{2k-1}\rangle+c_{2k}|\phi_{2k}\rangle)
+c_{2n+1}|\phi_{2n+1}\rangle$, where the basis ket is given by
$\{|s_{1z},N_{r1},s_{2z},N_{r2},$ $\cdots
,s_{nz},N_{rn},s_{az}\rangle\}$ as before. For
$|\phi_{2k-1}\rangle$ state $N_{rj}=0$ and $s_{az}=s_{jz}=-1/2$
except $s_{kz}=1/2$, for $|\phi_{2k}\rangle$ state $N_{rj}=0$
except $N_{rk}=1$ and  $s_{az}=s_{jz}=-1/2$, and for
$|\phi_{2n+1}\rangle$ state $N_{rj}=0$, $s_{jz}=-1/2$ and
$s_{az}=1/2$, where $1\leq j \leq n$.
Here,  the coefficients are given by
\begin{eqnarray}
\label{ck}
c_{2k-1}\!\!\!\!&=&\!\!\!\!\frac{1}{\sqrt{n}q^2}[(n\theta^2\!\!+\!\!\phi^2\!\!\cos q)\!\cos\!\!\sqrt{n}\eta\!
-\!2\sqrt{n}\phi\theta\!\sin^2\!\frac{q}{2}\!\sin\!\!\sqrt{n}\eta],\nonumber\\
c_{2k}\!\!\!\!&=&\!\!\!\!-\frac{\sin q}{\sqrt{n}q}(\phi\cos\sqrt{n}\eta+\sqrt{n}\theta\sin\sqrt{n}\eta),\\
c_{2n+1}\!\!\!\!&=&\!\!\!\frac{1}{q^2}[(\phi^2\!+\!n\theta^2\!\cos q)\!\sin\!\sqrt{n}\eta-2\sqrt{n}\phi\theta\!
\sin^2\!\frac{q}{2}\!\cos\!\sqrt{n}\eta].\nonumber
\end{eqnarray}

\begin{figure}[t]
\vspace{-0.5cm}
\includegraphics[width=0.45\textwidth]{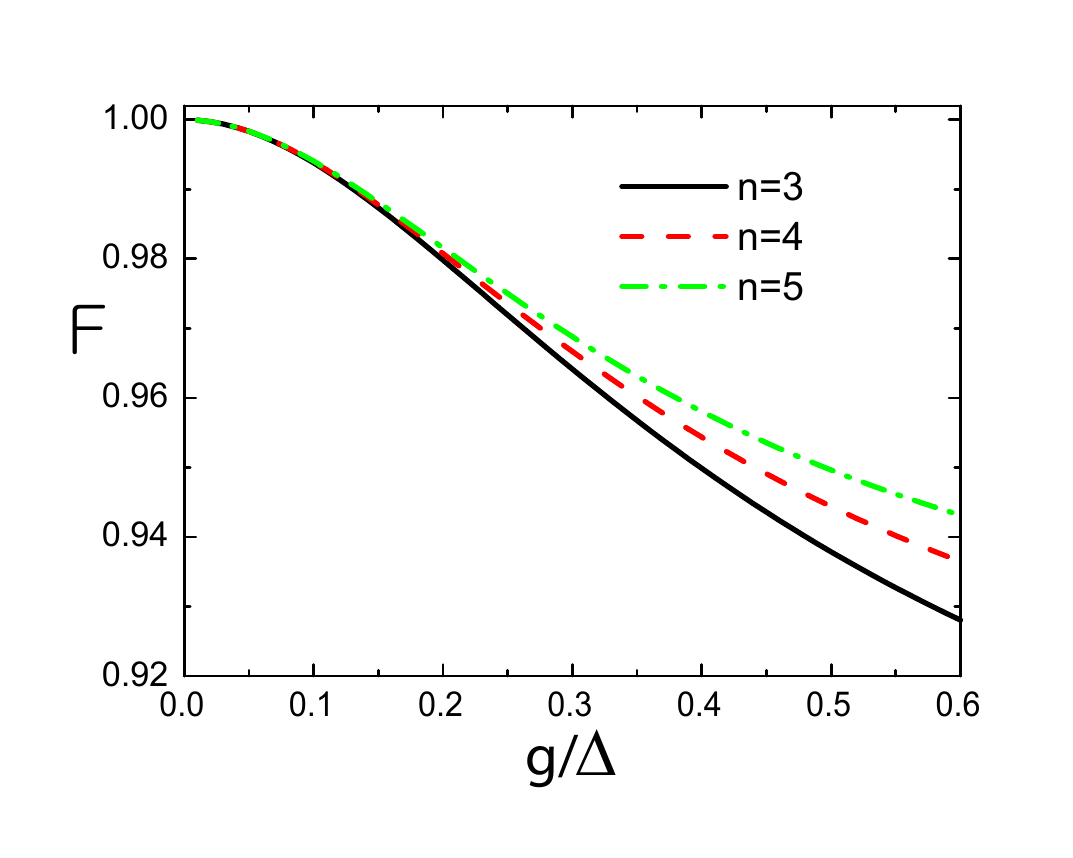}
\vspace{0cm}
\caption{Fidelity for W state formation by using the design in Fig. \ref{fig1}(b)
as a function of qubit-resonator coupling strength for $n=3,4,5$.}
\label{fig4}
\vspace{0cm}
\end{figure}

The fidelity for the W state formation can be given by
$F=|\langle\psi_W| \psi'_W\rangle|
=(1/\sqrt{n})\sum^n_{k=1}c_{2k-1}=\sqrt{n}{\cal C}$  with ${\cal
C}=c_{2k-1}$, where  $|\psi_W\rangle$ is the W state in the
original basis. For a tripartite system with $n=3$, for example,
$|\psi_W\rangle$ is explicitly written as
$|\psi_W\rangle=(1/\sqrt{3})(|\frac12,0,-\frac12,0,-\frac12,0,-\frac12
\rangle + |-\frac12,0, \frac12,0,$ $-\frac12,0,-\frac12  \rangle +
|-\frac12,0,-\frac12,0,\frac12,0,-\frac12 \rangle)$. In
superconducting qubit systems the three-qubit W states
\cite{WKim,Mlynek} have been demonstrated in experiments by using
sequential gates \cite{Neeley,Altomare,Sun}. In Fig. \ref{fig4}
the fidelities $F$  are close to 1 for weak coupling, but in this
case the gap of energy levels between the W state and the other
states become too small. On the other hand, as the coupling
increases, the energy gap increases but fidelity decreases. The
fidelity increases slightly along with $n$ due to the reduced
occupation probability of the bridge qubit state for higher $n$.

The present scheme suffers from the decoherence problem.
It is known that decoherence rate increases with the system size \cite{Unruh,Palma}.
In  Fig. \ref{fig1}(a) two qubits and a bridge qubit are coupled to
two resonators, which may cause decoherence due to hybridization of the qubits and the cavity.
This decoherence effect can be estimated by obtaining the eigenstates of the hybridized system.

From the symmetric eigenstate $|\psi^*\rangle_s=(1/\sqrt{2})(1~1~0)^T$
of the Hamiltonian $H^*_{2,2}$ in Eq. (\ref{H22*}) for $n=2$
we obtain the ground state with eigenvalue $\varepsilon^{q1}_2$ of the Hamiltonian $H_2$ in original basis such as
$|\psi\rangle_g=c_1|\frac12,0,-\frac12,0,-\frac12\rangle+c_2|-\frac12,1,-\frac12,0,-\frac12\rangle
+c_3|-\frac12,0,\frac12,0,-\frac12\rangle+c_4|-\frac12,0,-\frac12,1,-\frac12\rangle
+c_5|-\frac12,0,-\frac12,0,\frac12\rangle$, where $c_i$'s are given by Eq. (\ref{ck})
with $n=2$.
Also, from the antisymmetric  eigenstate $|\psi^*\rangle_a=(1/\sqrt{2})(1~-1~0)^T$
we have the excited state with eigenvalue $\varepsilon^{q2}_2$ such as
\begin{eqnarray}
|\psi\rangle_e=\frac{\cos\phi}{\sqrt{2}}|\frac12,0,-\frac12,0,-\frac12\rangle
-\frac{\sin\phi}{\sqrt{2}}|-\frac12,1,-\frac12,0,-\frac12\rangle\nonumber\\
-\frac{\cos\phi}{\sqrt{2}}|-\frac12,0,\frac12,0,-\frac12\rangle
+\frac{\sin\phi}{\sqrt{2}}|-\frac12,0,-\frac12,1,-\frac12\rangle.~~
\end{eqnarray}

\begin{figure}[t]
\vspace{-2.5cm}
\includegraphics[width=0.4\textwidth]{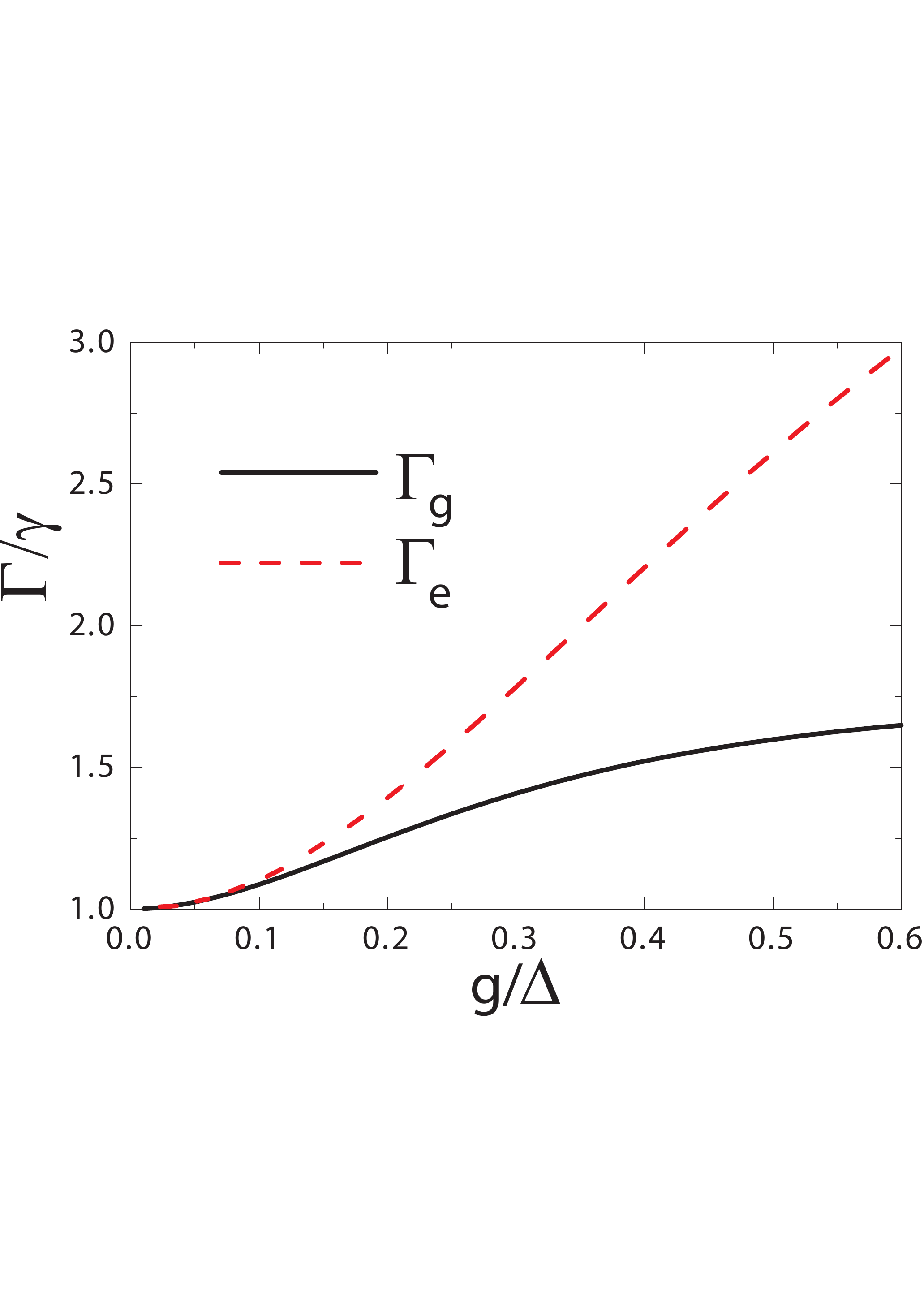}
\vspace{-2cm}
\caption{Relaxation rates of the ground and excited states of coupled qubit-resonator system
in Fig. \ref{fig1}(a) as a function of qubit-resonator coupling strength.
Here we set  $\kappa=12\gamma$ \cite{Blais1} and $\gamma'=\gamma$.}
\label{fig5}
\vspace{0cm}
\end{figure}

From the eigenstate in original basis we can estimate the decay rate
of the ground(excited) state $\Gamma_{g(e)}$ \cite{Blais1} as
\begin{eqnarray}
\Gamma_g&=&(c^2_1+c^3_3)\gamma+(c^2_2+c^2_4)\kappa+c^2_5\gamma',\\
\Gamma_e&=&\gamma\cos^2\phi+\kappa\sin^2\phi,\nonumber
\end{eqnarray}
where $\gamma$ and $\gamma'$ are the relaxation rate of qubit and
bridge qubit, respectively, and $\kappa$ that of cavity. Due to
coupling to the cavity with large relaxation rate $\kappa>\gamma$
the qubit state decays by the Purcell effect. In Fig. \ref{fig5}
we observe that the relaxation rate of eigenstates increases along
with the qubit-resonator coupling $g$. The ratio of relaxation
rate of the coupled system to that of the uncoupled qubit remains
in the same order for reasonable coupling $g$ so that the present
scheme may be viable in real experiments.



In summary, a coupling scheme for qubits in circuit-QED
architecture with a bridge qubit introduced between cavities is
studied. The cavity modes interacting with the bridge qubit
mediate the coupling between qubits. We derive a transformation
producing an exact representation of two-qubit coupling for
arbitrary number of qubits. Our active coupling scheme enables
switching function and control of coupling between  qubits. For
$n$ qubits coupled with each other the system can produce the W
state as an eigenstate of the system. We find that the coupled
two-qubit system will be viable in experiments.

{\it Acknowledgments.}$-$ {This research was supported by Basic
Science Research Program through the National Research Foundation
of Korea (NRF) funded by the Ministry of Education, Science and
Technology (2011-0023467) and by the IT R$\&$D program of
MOTIE/KEIT [10043464(2012)]}.

\appendix
\section{Transformation Matrix}

The transformation can be explicitly evaluated for coupled $n$
qubits. Here we, for simplicity, consider identical qubits and
resonators, $\phi_j=\phi$ and $\theta_j=\theta$ and thus
\begin{eqnarray}
U_{n}\!=\!e^{M}\!=\!e^{-\sum^n_{j=1}[\phi(a^\dagger_j\sigma_{j-}-a_j\sigma_{j+})
+\theta(a^\dagger_j\sigma_{a-}-a_j\sigma_{a+})]}.
\end{eqnarray}
Then, the $U_{n}$ can be represented as a block-partitioned
$(2n+1)\times(2n+1)$ matrix consisting of $Q,T,C$ and $C'$  for
${\cal N}_e=1$ as shown in Eq. (\ref{MUn}). If we expand $U_{n}$
by using the relation,
$e^{M}=1+M+\frac{1}{2!}M^2+\frac{1}{3!}M^3+\cdot\cdot\cdot$, we
obtain
\begin{eqnarray}
Q_{11}&=&1+\phi^2\sum^\infty_{m=1}\frac{(-1)^m}{(2m)!}a_m \\
Q_{22}&=&\sum^\infty_{m=1}\frac{(-1)^{m-1}}{(2m-2)!}a_{m}\\
Q_{12}&=&\phi\sum^\infty_{m=1}\frac{(-1)^{m-1}}{(2m-1)!}a_{m}\\
Q_{21}&=&-Q_{12}
\end{eqnarray}
with
\begin{eqnarray}
a_1&=&1,\nonumber\\
a_2&=&\phi^2+\theta^2, \nonumber\\
a_3&=&\phi^4+2\phi^2\theta^2+n\theta^4, \\
a_4&=&\phi^6+3\phi^4\theta^2+3n\phi^2\theta^4+n^2\theta^6,\nonumber\\
\cdots && \nonumber
\end{eqnarray}
This series has the following general form
\begin{eqnarray}
a_m&=&q^{2(m-1)}-(n-1)\theta^2\sum^{m-1}_{l=1}q^{2(m-l-1)}\phi^{2(l-1)}\nonumber\\
&=&\frac{1}{n}q^{2(m-1)}+\frac{n-1}{n}\phi^{2(m-1)}
\end{eqnarray}
with $q=\sqrt{\phi^2+n\theta^2}$ and thus the matrix $Q$ can be written as
\begin{eqnarray}
Q=\left(
\begin{array}{cc}
\frac{\phi^2}{nq^2}\cos q+\frac{n-1}{n}\cos\phi+\frac{\theta^2}{q^2} &
\frac{\phi}{nq}\sin q+\frac{n-1}{n}\sin\phi \\
-\frac{\phi}{nq}\sin q-\frac{n-1}{n}\sin\phi &
\frac{1}{n}\cos q+\frac{n-1}{n}\cos\phi\\
\end{array}
\right).\nonumber\\
\end{eqnarray}

For the matrix $T$ we have similar results such that
\begin{eqnarray}
T_{11}&=&\phi^2\sum^\infty_{m=1}\frac{(-1)^{m+1}}{(2m+2)!}b_m \\
T_{22}&=&\sum^\infty_{m=1}\frac{(-1)^m}{(2m)!}b_m\\
T_{12}&=&\phi\sum^\infty_{m=1}\frac{(-1)^m}{(2m+1)!}b_m\\
T_{21}&=&-T_{12}
\end{eqnarray}
with
\begin{eqnarray}
b_1&=&\theta^2,\nonumber\\
b_2&=&\theta^2(2\phi^2+n\theta^2),\nonumber\\
b_3&=&\theta^2(3\phi^4+3n\phi^2\theta^2+n^2\theta^4), \\
b_4&=&\theta^2(4\phi^6+6n\phi^4\theta^2+4n^2\phi^2\theta^4+n^3\theta^6),\nonumber\\
\cdots &&\nonumber
\end{eqnarray}
and thus
\begin{eqnarray}
b_m&=&\frac{1}{n}\left(q^{2m}-\phi^{2m}\right),
\end{eqnarray}
resulting in
\begin{eqnarray}
T=\left(
\begin{array}{cc}
\frac{\phi^2}{nq^2}\cos q-\frac{1}{n}\cos\phi+\frac{\theta^2}{q^2} &
\frac{\phi}{nq}\sin q-\frac{1}{n}\sin\phi \\
-\frac{\phi}{nq}\sin q+\frac{1}{n}\sin\phi &
\frac{1}{n}\cos q-\frac{1}{n}\cos\phi\\
\end{array}
\right).\nonumber\\
\end{eqnarray}

The elements of matrices $C$ and $C'$ also can be evaluated similarly as
\begin{eqnarray}
C_1&=&-\frac{1}{2!}\phi\theta+\frac{1}{4!}\phi\theta(\phi^2+n\theta^2)
-\frac{1}{6!}\phi\theta(\phi^2+n\theta^2)^2+\cdots\nonumber \\
\\
C_2&=&-\theta+\frac{1}{3!}\theta(\phi^2+n\theta^2)-\frac{1}{5!}\theta(\phi^2+n\theta^2)^2
+\cdots\nonumber\\
\end{eqnarray}
$C'_1=C_1$, and $C'_2=-C_2$,
resulting in
\begin{eqnarray}
C&=&\left(
\begin{array}{c}
\frac{\phi\theta}{q^2}(\cos q-1)\\
-\frac{\theta}{q}\sin q \\
\end{array}
\right),\\
C'&=&\left(
\begin{array}{cc}
\frac{\phi\theta}{q^2}(\cos q-1) &
\frac{\theta}{q}\sin q \\
\end{array}
\right).
\end{eqnarray}
Finally, the matrix element, $[U_{n}]_{2n+1,2n+1}$, can be evaluated as
\begin{eqnarray}
[U_{n}]_{2n+1,2n+1}&=&1-\frac{1}{2!}n\theta^2+\frac{1}{4!}n\theta^2(\phi^2+n\theta^2)
-\cdots\nonumber\\
&=&\frac{\phi^2+n\theta^2\cos q}{q^2}.
\end{eqnarray}




\end{document}